%% file: M-MIMO_and_Delay_spread_ArXiv.tex
\documentclass[12pt,draftcls,onecolumn]{IEEEtran}

\usepackage[english]{babel}
\usepackage[T1]{fontenc} 
\usepackage{mathrsfs} 
\usepackage{verbatim}
\usepackage{comment}

\usepackage{latexsym, amssymb, dsfont, color, url}
\usepackage[cmex10]{amsmath}  
\usepackage{graphicx}
\usepackage{epsfig}
\usepackage{float}
\usepackage{dblfloatfix}
\usepackage[caption=false,font=footnotesize]{subfig} 

\usepackage{array}    
\usepackage{tabulary} 
\usepackage{multirow} 

\usepackage{cite}
\usepackage{cases}    
\usepackage{url}
\usepackage{balance}

\DeclareGraphicsExtensions{.eps,.pdf,.png,.jpg,.gif,.jpeg,.pstex}

\input{MyMacros.tex}

\newcounter{MYcounter}

\newcommand{\insertonefig}[5]
{
\begin{figure}[!ht]
\begin{center}
\includegraphics[angle=#1,width=#2,clip,keepaspectratio]{#3}
\end{center}
\caption{#5}
{#4}
\end{figure}
}

\newcommand{\inserttwofigsV}[6]
{
\begin{figure}[!ht]
\begin{center}$
\begin{array}{c}
\includegraphics[angle=#1,width=#2,clip,keepaspectratio]{#3} \\
\text{(a)} \\
\linebreak[4] \\
\includegraphics[angle=#1,width=#2,clip,keepaspectratio]{#4} \\
\text{(b)} \\
\end{array}$
\end{center}
\caption{#6}
{#5}
\end{figure}
}

\begin{document}


\title{The Effect of Diversity Combining on ISI\\in Massive MIMO}


\author{
\IEEEauthorblockN{Arkady~Molev~Shteiman, Stefano~Galli, Laurence~Mailaender, Xiao~Feng~Qi}
\IEEEauthorblockA{Radio Algorithms Research\\NJ Research Center\\Futurewei Technologies, Inc.\\Bridgewater, New Jersey, USA\\ \{Arkady.Molev.Shteiman,Stefano.Galli,Laurence.Mailaender,Xiao.Feng.Qi\}@huawei.com}
}


\maketitle

\begin{abstract}
We analyze the resiliency of Massive Multiple-Input Multiple-Output (M-MIMO) systems to Inter-Symbol Interference (ISI) when diversity combining techniques are used at the Base Station (BS). We show that Maximum Ratio Combining (MRC) alone can equalize an ISI channel as the number of antennas grows unbounded. Additional constraints on the nature of the channel must be postulated depending on whether the information of the Angle-of-Arrival (AoA) is exploited at the receiver or not. 
Interestingly, the simpler Equal Gain Combiner (EGC) receiver is also able to equalize the channel as the number of antennas grows but, in this case, at least one channel path must be Ricean faded. These findings are confirmed via simulation on WSSUS channels and channels generated with a ray tracing engine simulating a real BS deployment in downtown Hong Kong and Shanghai. Finally, the observed scaling law indicates that normalized ISI power decreases $N$-fold for every $N$-fold increase in the number of antennas at the BS. 
\end{abstract}

\thispagestyle{empty}


\section{Introduction}
M-MIMO systems use an excess of antennas at the BS to offer multiplexing gains in both the uplink and downlink and allocate the whole bandwidth to all single-antenna users simultaneously, while using simple (linear) processing \cite{Marzetta2010MassiveMIMO}. M-MIMO coherently processes the signals impinging on the array with receive combining in the uplink to separate the signals of different users, and then uses transmit precoding to perform fine spatial discrimination in the downlink. M-MIMO has received considerable attention in the past decade and 3GPP has included M-MIMO support in Release 13 (16 antennas) and 14 (32 antennas) for LTE, and release 15 (64 antennas) for New Radio (NR)/5G \cite{Wong2017key5G, OnggosanusiRahman2018}.

The use of higher frequencies allows deploying larger arrays which can provide gains to reduce higher path loss and provide multiplexing gains, at the cost of higher cell density and increased backhauling needs \cite{Jaber2016-5Gbackhaul,GalliJu2018baremetal}. However, regardless of the frequencies at which it is used, the degrees of freedom offered by M-MIMO provide robustness to many impairments such as hardware imperfections, phase noise, non-linearities, quantization errors, noise amplification, and inter-carrier interference \cite{YangMarzetta2013TXbeamforming,NgoLarsson2013EnergySpectralEfficiency, BjornsonDebbah2015HardwareScaling}.

Robustness of M-MIMO to ISI, on the other hand, appears to have been a less investigated topic while recent measurement results seem to indicate that ISI indeed decreases as the number of antennas grows \cite{PayamiTufvession2013measurementsMMIMO}. If indeed M-MIMO allows for the reduction of ISI not only in the downlink (transmit beamforming) but also in the uplink (receive combining), then the case for using multicarrier schemes (ISI mitigation techniques) would be weaker when compared to single carrier schemes. In fact, although OFDM is an attractive and popular choice for transmission schemes, it also exhibits several drawbacks such as transmission and power inefficiency, high peak-to-average ratio, and phase-noise sensitivity at high frequencies \cite{Pitarokoilis2012OptimalitySingleCarrier, WuWubben2016OFDNvsSC-FDE, AminjavaheriFarhang2017OFDMWithoutCP, BuzziColavolpe2018single}. These drawbacks may entail a price that is too high to pay if there is no need to equalize the channel because ISI vanishes anyway as the number of antennas grows. This topic has practical relevance because the choice of waveform in 3GPP is still open above 60~GHz, while below 60~GHz 3GPP has already agreed to multicarrier.

In this paper, we address the topic of the robustness of M-MIMO to ISI when the BS is equipped with linear combining techniques such as MRC or EGC and does not perform equalization. We show analytically and by computer simulations that ISI decreases asymptotically to zero as the number of BS antennas grows to infinity. More specifically, if AoA information is not exploited at the receiver, then it is necessary to operate in a rich scattering environment and assume Wide-Sense Stationary Uncorrelated Scattering (WSSUS) channels to get vanishing ISI via linear combining. However, if AoA information is exploited at the receiver, then the rich scattering assumption is not necessary anymore as it suffices to assume that each observable path has a different AoA and that the number of paths is smaller than the number of antennas.

Furthermore, we also report that the normalized ISI power decreases $N$-fold for every $N$-fold increase in the number of receive antennas, and that the same asymptotic decrease can be observed also when the receiver is equipped with a simple EGC receiver as long as the multipath channel has at least one path that is Rice distributed. 

\section{The System Model}  \label{sec.Modeling}
We consider here two types of scenarios, one characterized by low scattering and one by rich scattering, and their associated observation models. The corresponding channel model here considered is a wideband, time-flat, and frequency-selective SIMO channel model between a single antenna node and a Base Station equipped with $M$ antennas (Uniform Linear Array , ULA). A wideband channel model with spatial structure is derived in Appendix \ref{App.WidebandChannel}.

\subsection{Low scattering scenario}
We consider here the model in Appendix \ref{App.WidebandChannel} for a finite number of paths $P$. In this case, the $\medmuskip=0muM\times 1$ observation vector at sampling time $k$ is:
\begin{eqnarray}
  \by[k] &=& \sum_{n=0}^{L-1} \bh[n] x[k-n]+ \bn[k]\\
         &=& \bH \bx[k] + \bn[k],    \label{eq.observation}
\end{eqnarray}
where $\bh[n]$ is the vector in \eqref{eq.APPfinal} collecting the channel taps at lag $n$ of the ISI channels experienced by the $M$ BS antennas; $\bH=[\bh[0],\ldots,\bh[L-1]]$ is the $\medmuskip=0mu M\times L$ channel matrix; $\bx[k]=[x[k],\ldots,x[k-L+1]]^T$ is the channel state vector collecting the last $L$ transmitted symbols; $\bn[k]$ is the $\medmuskip=0mu M\times 1$ noise vector with jointly Gaussian, zero mean, and complex components; and $L$ is the duration of all equivalent discrete-time ISI channels which are here assumed to be all the same for simplicity.

\subsection{Rich scattering scenario}
For this scenario, the observation model on the $m$-th diversity branch can be expressed as $(1\leq m\leq M)$:
\begin{equation}\label{eq.SIMO-ISI-observation-branch}
  \by_m = \bH_m \bs + \bn_m,
\end{equation}
where $\by_m$ is the vector of $(\medmuskip=0mu N+L_i-1){\times}1$ received samples, $\bs$ is the $\medmuskip=0mu N\times 1$ vector of transmitted symbols, $\bH_m$ is a $\medmuskip=0mu (N+L-1)\times N$ random channel matrix with elements that are i.i.d, zero mean, complex, and Gaussian random variables, and $\bn_i$ is the random noise vector.

By stacking vectors and matrices, we can rewrite the observation model \eqref{eq.SIMO-ISI-observation-branch} as:
\begin{equation}\label{eq.SIMO-ISI-observation}
  \by = \bH \bx + \bn,
\end{equation}
having posed:
%
\begin{equation}
    \by = \begin{bmatrix}
           \by_1 \\
           \by_2 \\
           \vdots \\
           \by_M
         \end{bmatrix}
\text{,~~~~~~}
    \bn = \begin{bmatrix}
           \bn_1 \\
           \bn_2 \\
           \vdots \\
           \bn_M
         \end{bmatrix}
\text{,~~~and~~~}
    \bH = \begin{bmatrix}
           \bH_1 \\
           \bH_2 \\
           \vdots \\
           \bH_M
         \end{bmatrix}.
\end{equation}

In the above equation, $\bH$ is an $\medmuskip=0mu M(N+L-1) \times N$ matrix, while $\by$ and $\bn$ are $\medmuskip=0mu M(N+L-1) \times 1$ vectors.


\section{Maximum Ratio Combiner (MRC) for Frequency Selective Fading}  \label{sec.SIMO-MRC}
The optimal Maximum Likelihood (ML) decision rule for the model in eq. \eqref{eq.observation} or \eqref{eq.SIMO-ISI-observation} can be expressed as follows:
\begin{IEEEeqnarray}{rCl}
  \tilde{\bx} &=& \argmax_{\bx} ~\{p(\bx \mid \by)\} \nonumber \\
              &=& \argmin_{\bx} ~\{\parallel \by - \bH \bx \parallel ^2\}  \nonumber \\
              &=& \argmin_{\bx} ~\{\bx^\herm \bH^\herm \bH \bx -2 \bx^\herm \bH^\herm \by \}.  \label{eq.SISO-ISI-MAPvec}
\end{IEEEeqnarray}

Equation \eqref{eq.SISO-ISI-MAPvec} shows that sequence $\by$ is actually irrelevant to optimal detection once the $\medmuskip=0mu L\times 1$ vector $\br=\bH^\herm \by$ is available. Thus, $\br$ is a sufficient statistic for ML detection and can be calculated as follows depending on the considered observation model. For example, using \eqref{eq.SIMO-ISI-observation} we obtain:
\begin{eqnarray}
  \br^{\textrm{(MRC)}}    &=& \bH^\herm\by \nonumber\\
                          &=& \bM \bs + \bv[k],  \label{eq.SIMO-MRC-RichScatter}
\end{eqnarray}
where $\bM=\bH^\herm\bH$ is the Gramian of $\bH$ and channel $\bH$ is defined as in model \eqref{eq.observation} or \eqref{eq.SIMO-ISI-observation}, respecitvely. The non-white noise term $\bv=\bH^\herm \bn$ has a covariance matrix equal to $\bR_\bv = E\{\bv \bv^\herm \} = \bM \sigma^2$. This is the classic MRC receiver where $\bH^\herm$ acts as a spatio-temporal Matched Filter (MF).

\section{Equal Gain Combining (EGC)} \label{sec.SIMO-EGC-ISI}
Another classical combining technique is Equal Gain Combining (EGC), which is a simpler coherent technique than MRC. In EGC, signals on all diversity branches are co-phased to provide coherent signal addition but then the same gain is applied to all diversity branches. Thus, in EGC, channel amplitude estimation is not required.

Using for example model \eqref{eq.SIMO-ISI-observation}, the observation model for the case that all diversity branch gains are equal to $1/M$ becomes:
\begin{equation}\label{eq.SIMO-ISI-EGC}
  \br^{\textrm{(EGC)}} = \frac{1}{M} \sum_{i=1}^{M} \by_i = \overline{\bH} \bx + \overline{\bn},
\end{equation}
where the matrix elements $\{\overline{\bH}\}_{k,l}=\frac{1}{M}\sum_{i=1}^{M}\{\bH_i\}_{k,l}$ are the sample means of the channel taps across antennas.

The observation model in eq. \eqref{eq.SIMO-ISI-EGC} suggests that zero mean channel taps would vanish as $M$ grows unbounded since, by the law of large numbers, the sample means would converge almost surely to the actual means. 

\section{The Effect of Large Arrays on ISI} \label{sec.Effectanalytical}
If the BS is ``looking'' towards direction $\alpha_{\textit{BS}}$, the amount of ISI that the BS would experience is inversely proportional to the directional selectivity of the receiver antenna array. This can be quantified by calculating the correlation $\zeta^{(M)}_n$ between the $n$-th channel tap across antennas ($\bh[n]$ in \eqref{eq.observation}) and the receive beamforming steering vector $\ba(\alpha_{\textit{BS}})$ (see Appendix \ref{App.WidebandChannel}) of the BS in the direction of angle $\alpha_{\textit{BS}}$:
\begin{IEEEeqnarray}{rCl}\label{eq.SIMO-ISI-corr}
  \zeta^{(M)}_n &=& \frac{\bh[n]^\herm \ba(\alpha_{\textit{BS}})}{\parallel\bh[n]\parallel \parallel \ba(\alpha_{\textit{BS}})\parallel} \nonumber \\
  &=& \frac{\sum_{j=1}^{P} g^*_{j,n} \ba^\herm(\alpha_{j}) \ba(\alpha_{\textit{BS}})}{\parallel\bh[n]\parallel \parallel \ba(\alpha_{\textit{BS}})\parallel}.
\end{IEEEeqnarray}

As shown in Appendix \ref{App.Proof}, we can prove the following asymptotic result for any tap $n$ under the assumption that the Angles of Arrival (AoAs) of the multipaths are all different:
\begin{equation}\label{eq.SIMO-ISI-rholimit}
\lim_{M \to \infty} \zeta^{(M)}_n =
  \begin{cases}
        \frac{c_k}{\sqrt{\sum_{i=1}^{P}|c_i|^2}} & \text{, if  $\alpha_{\textit{BS}}=\alpha_k$, $n=0$}
        \\
        0 & \text{, otherwise}
  \end{cases}
  .
\end{equation}

This result confirms that, when the BS performs receive beamforming in a direction $\alpha_{\textit{BS}}$ equal to one of the unique multipath's AoAs, then only that single path survives yielding an ISI free observation with zero RMS-DS. The result holds regardless of any statistical assumption made on the paths' fading coefficients and holds for instantaneous channel vectors.

In some (rare) cases, two or more signal paths may arrive at the BS array with the same AoA but different delays. If the difference between these same-angle delays is much smaller than the reciprocal of the transmit/receive filters' bandwidth, then no additional ISI is created because the samples of $\tilde{p}(t)$ at the various sampling times would be nearly zero (see also Appendix \ref{App.WidebandChannel}). On the other hand, if the time difference is considerable, then residual ISI may be present after combining and regardless of the number of antennas at the BS.
\insertonefig   {0}{8.8cm}{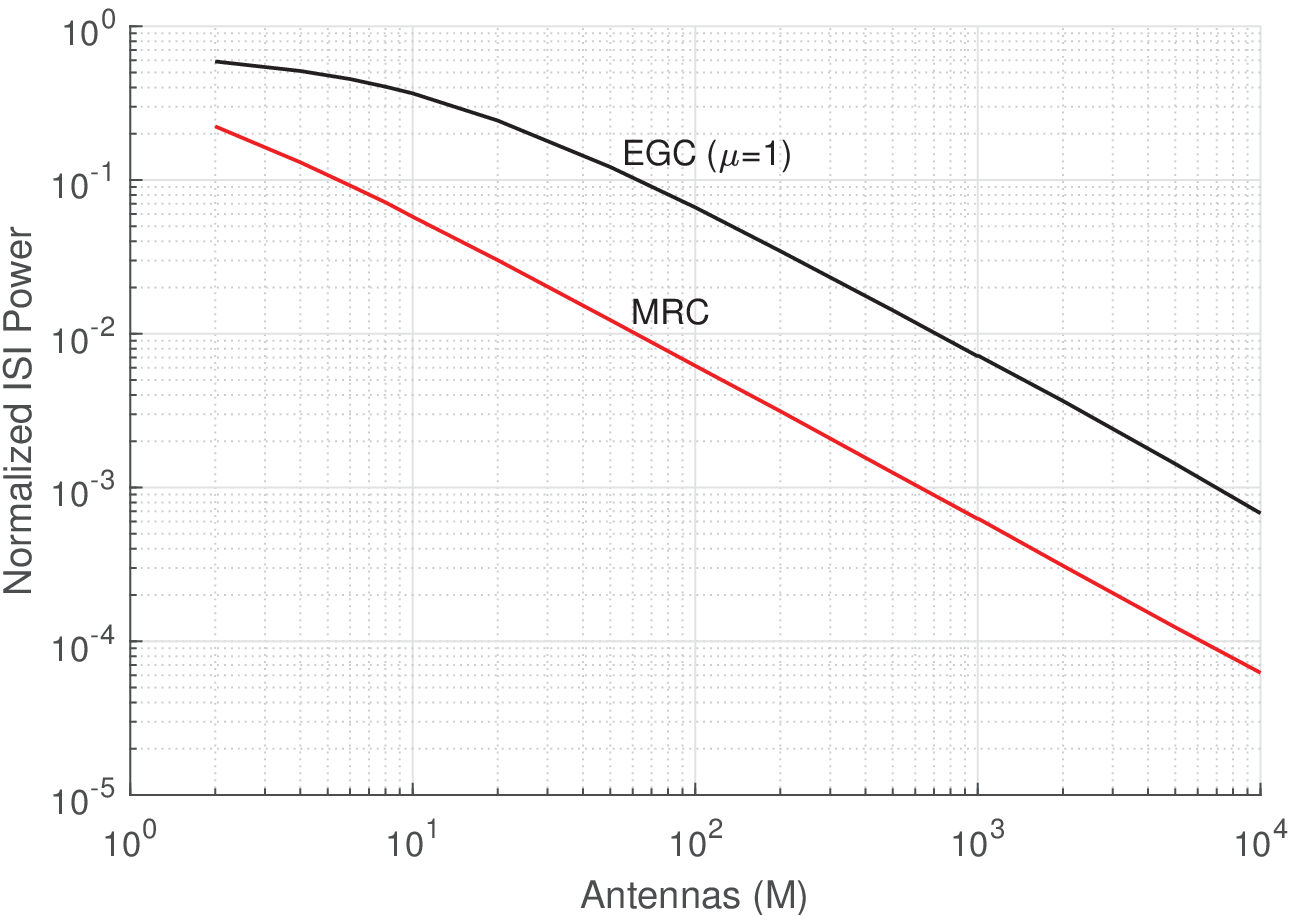}{\label{fig.ISIvsAntennas}}
{Log-Log plot of normalized ISI power $\rho$ vs number of Antennas $M$ for MRC and EGC receivers. One thousand WSSUS Rayleigh and Rice channels were generated.}

Let us now consider the case when the receiver does not exploit the knowledge of the AoA of multipath or ignores the underlying spatial structure of the channel. The Gramian $\bM$ in the post-MRC observation has elements given by $\bM_{(m,n)}=\bh^\herm[m] \bh[n]$. Proving that MRC eliminates ISI asymptotically is equivalent to proving that the Gramian becomes diagonal as the number of BS antennas goes to infinity - or, equivalently, that the normalized off-diagonal elements $D^{(M)}_{m,n} = \frac{\bg^\herm_m A^\herm A  \bg_n} {\| \bh[m] \| \| \bh[n] \|} $ vanish. Following the same tedious algebra used for proving eq. \eqref{eq.SIMO-ISI-rholimit}, we can write:
\begin{equation}
\begin{split}
  D_{m,n} &= \lim_{M \to \infty}   D^{(M)}_{m,n} \\
    &=  {
    \begin{cases}
        \frac{ \bg_m^\herm \bg_n}{\sqrt{\left(\bg_m^\herm \bg_m\right) \left(\bg_n^\herm \bg_n\right)}} & \text{, for $m \neq n$}
        \\
         1 & \text{, for $m = n$}
  \end{cases}.
}
\end{split}
\end{equation}

As shown above, an infinite number of antennas is not enough to ensure orthogonality between the instantaneous channel vectors with an MRC. However, taking the limit for $P \to \infty$  (rich scattering environment), then the $P$ channel gains become i.i.d, zero mean, complex Gaussian random variables (Rayleigh fading) thanks to the Central Limit Theorem. Finally, the numerator of $D_{m,n}$ vanishes thanks to the Law of Large numbers:
\begin{equation}\label{bla}
   \lim_{P \to \infty}  \frac{\bg_m^\herm \bg_n}{\sqrt{\left(\bg_m^\herm \bg_m\right) \left(\bg_n^\herm \bg_n\right)}} \xrightarrow{a.s.} \frac{E\{ g^*_{i,m} g_{i,n}\}}{E\{|g_{i,m}|^2\}} = 0
\end{equation}
%

\section{Simulation Results}
In the previous section it was shown that, for large arrays, instantaneous channel vectors are asymptotically orthogonal if they do not share a common angle. Furthermore, for large arrays and rich scattering, an MRC receiver diagonalizes the Gramian leading to an ISI-free observation after combining. In this Section, the analytical results will be verified by computer simulations under the assumption of perfect Channel State Information (CSI) at the receiver.

The following fading channels where simulated:
\begin{enumerate}
  \item WSSUS Rayleigh: complex Gaussian uncorrelated taps with zero mean and unitary power.
  \item WSSUS Rice: complex Gaussian uncorrelated taps where the first tap has non-zero mean $\mu>0$ (Line-of-Sight, LOS, component) while all other taps have zero mean and unitary power.
  \item Channels obtained with the Volcano ray-tracing tool for the cities of Hong Kong and Shanghai  \cite{Link:Volcano,Newbury2017Volcano}.
\end{enumerate}
\inserttwofigsV {0}{8.8cm}{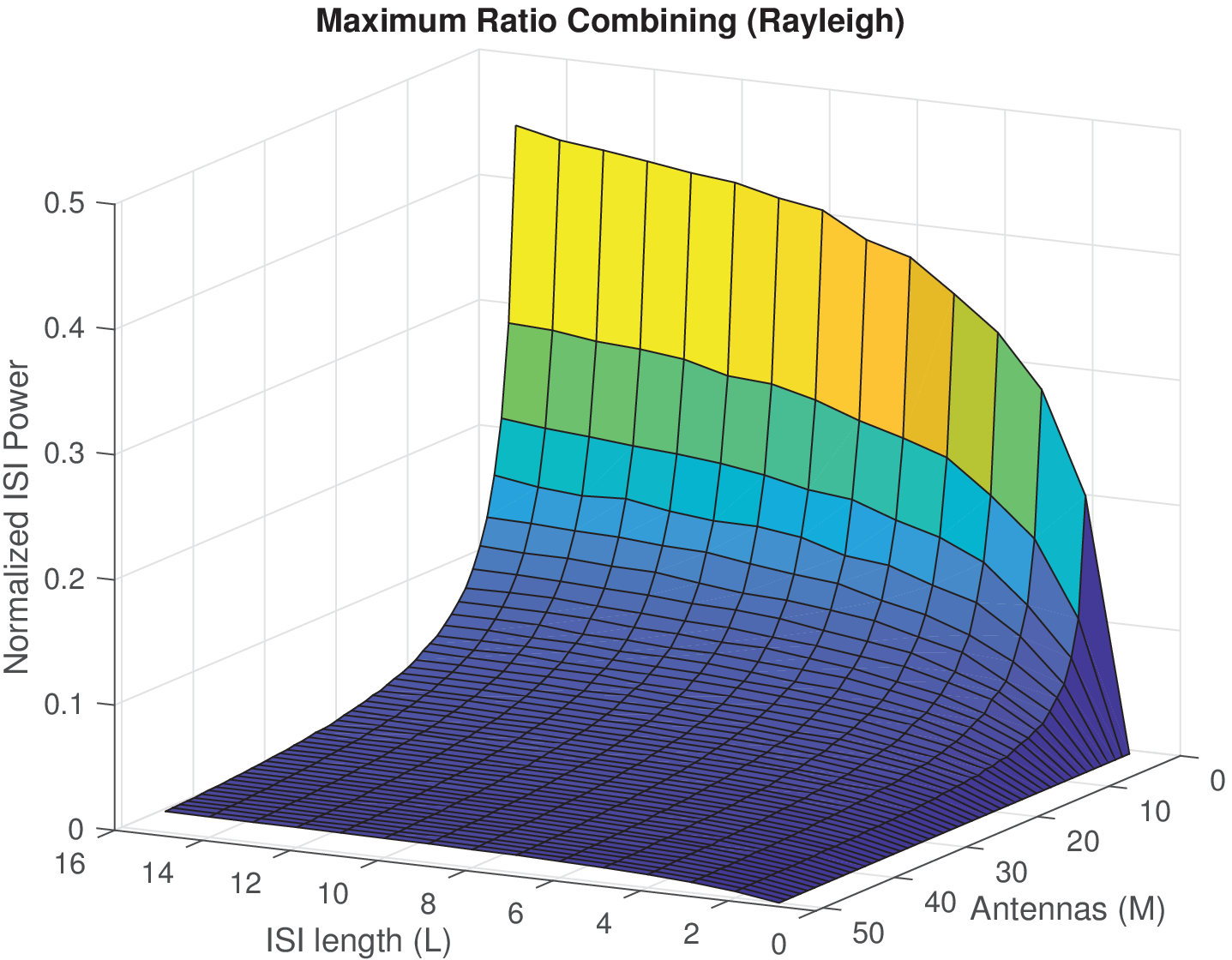}{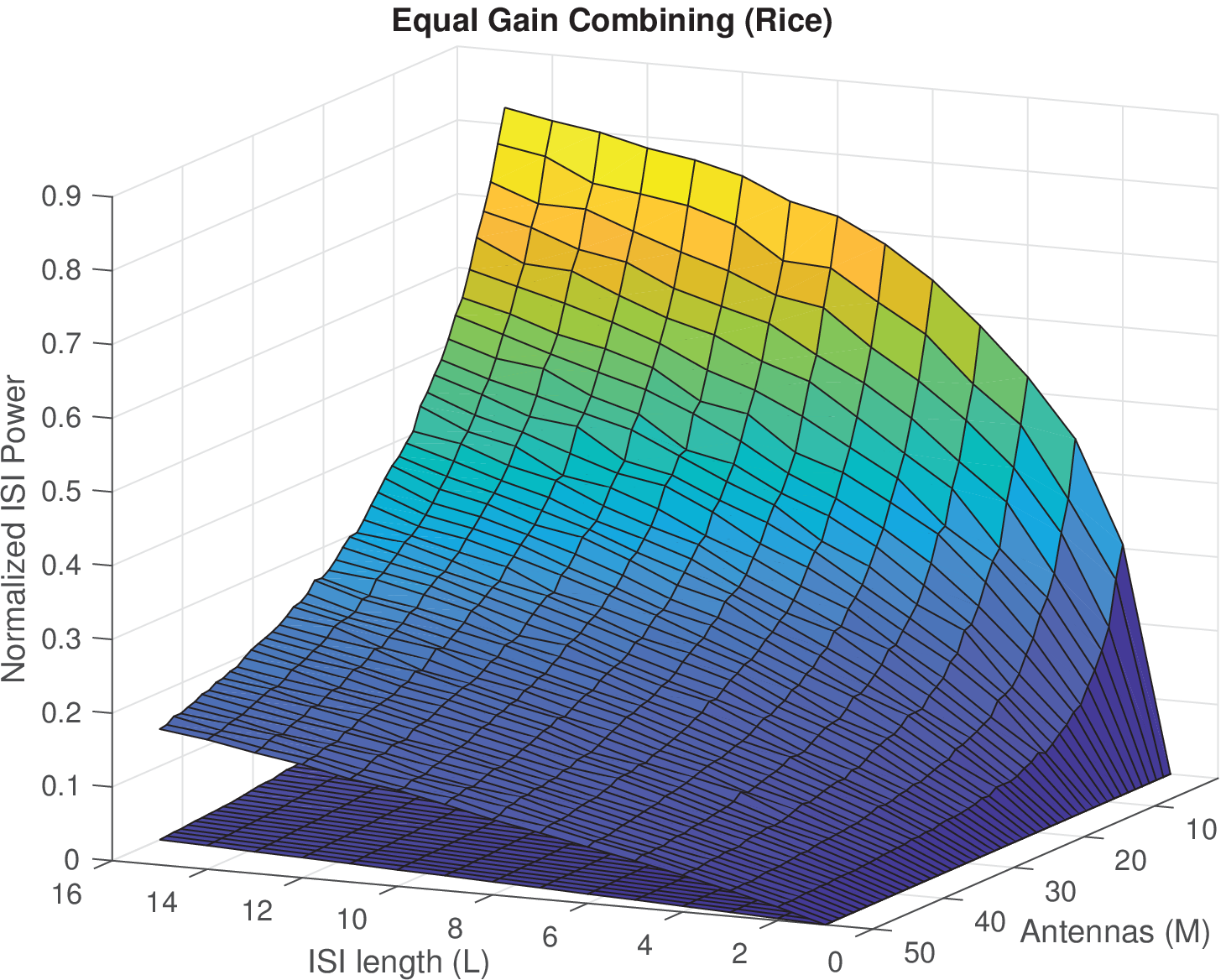}{\label{fig.MRCandEGC2D}}
{Normalized ISI power $\rho$ vs number of Antennas $M$ and ISI channel length $L$: (a) MRC and WSSUS Rayleigh; (b) EGC and WSSUS Rice for two values of the mean value $\mu$ of the first tap. One thousand WSSUS channel realizations were generated.}

A simple measure of the amount of ISI in a channel is the normalized power $\rho$ in the elements outside the main diagonal of the channel matrix. The value of $\rho$ will be computed via computer simulations by randomly generating thousands of complex channel realizations of length $L$ with Gaussian real and imaginary components of normalized energy. The normalized energy in the elements outside the main diagonal of the channel matrix takes the form below:
\begin{IEEEeqnarray}{rCl}
  \rho &=& \frac{\tr (\Xim^\herm \Xim)}{\tr (\Psim^\herm \Psim)} = \left(\frac{\|\Xim\|_F}{\|\Psim\|_F} \right)^{\!\!2}, \label{eq.ISIrho}
\end{IEEEeqnarray}
where $\Psim$ is a full channel matrix ($\bM$ or $\overline{\bH}$), $\Xim=\Psim - \diag\{\Psi_{1,1},~\Psi_{2,2},\ldots\}$ is the matrix containing only the off-diagonal elements of $\Psim$, and $\|\|_F$ denotes the Frobenius norm.

As Figure \ref{fig.ISIvsAntennas} confirms, the normalized ISI power decreases as the number of antennas grows when the receiver is equipped with either an MRC or an EGC. The scaling law we observe is that a $N$-fold increase in the number of antennas yields a $N$-fold reduction in ISI power. The decrease of ISI power for MRC receivers as the number of antennas grows holds regardless of the length of the ISI channel, as shown in Figure \ref{fig.MRCandEGC2D}.(a). This confirms the results reported in \cite{Eriksson1995}, but here no equalization is employed to combat ISI at the receiver. The simulation results reported in Figure \ref{fig.MRCandEGC2D}.(b) also confirm that an EGC yields vanishing ISI as the number of antennas grow and regardless of channel length $L$ for a Rice channel. In this case, the receiver does not exploit the knowledge of the AoAs of the multiple paths.

The Volcano ray tracing tool was used to generate channels based on an actual BS deployment in the two cities of Hong Kong and Shanghai. In both cases, a 3D map of the area was used, the carrier frequency was set to $2.1$~GHz, and all BSs were equipped with the same number of antennas (1, 16, 257, or 4096). For downtown Hong Kong, the considered area is $656$~m by $1052$~m covered by $57$ BSs. For Shanghai, the modeled area is the Jiao Tong University campus. The considered area has an extension of $1680$~m by $2176$~m, and is covered by $28$ BSs.

Given a BS and a uniformly distributed random user location in the 3D map, the Volcano tool generates the channels between the user and each antenna at the BS. The RMS-DS is calculated after diversity combining (perfect CSI) and the RMS-DS Cumulative Distribution Function (CDF) is obtained after collecting 100,000 values. In this case, the receiver exploits the knowledge of the AoAs of the multiple paths.

Figure \ref{fig.HongKongCDF} shows the obtained CDFs for the MRC and EGC cases. As the number of antennas grows, the ISI after combining decreases as predicted. We also point out that, for Hong Kong, the probability of encountering a LOS channel is only $20\%$ and, nevertheless, the EGC is effective anyway in reducing the channel RMS-DS. However, the rate of decrease of the RMS-DS as antennas grow is slower at high percentiles, especially for the EGC, thus confirming that for an arbitrarily large but finite number of antennas there may always be some residual ISI in the channel when only conjugate combining is performed. Similar results are obtained for the case of Shanghai, as shown in Figure \ref{fig.ShanghaiCDF}.

\section{Conclusions}
We have shown that, for WSSUS Rayleigh/Rice faded channels, ISI decreases for increasing receive diversity when either MRC or EGC are employed at the receiver and we have also reported that ISI power decreases $N$-fold for a $N$-fold increase in the number of antennas at the BS. 

Both MRC and EGC receivers are coherent receivers and carrier phase for co-phasing the signals across diversity branches is required. An MRC receiver also needs to estimate the channel tap amplitudes on each diversity branch and this channel estimation must be performed before experiencing any reduction in ISI. On the other hand, an EGC receiver does not need to estimate channel tap amplitudes and would experience ISI reduction before estimating the channel. 

If high spectral efficiency is not the prime design objective and radiated energy-efficiency is pursued (two design criteria suitable for mmW communications), then transmit conjugate beamforming and receive linear combining (MRC/EGC) are very attractive and simple solutions together with the use of single-carrier transmission schemes.

\inserttwofigsV {0}{7.5cm}{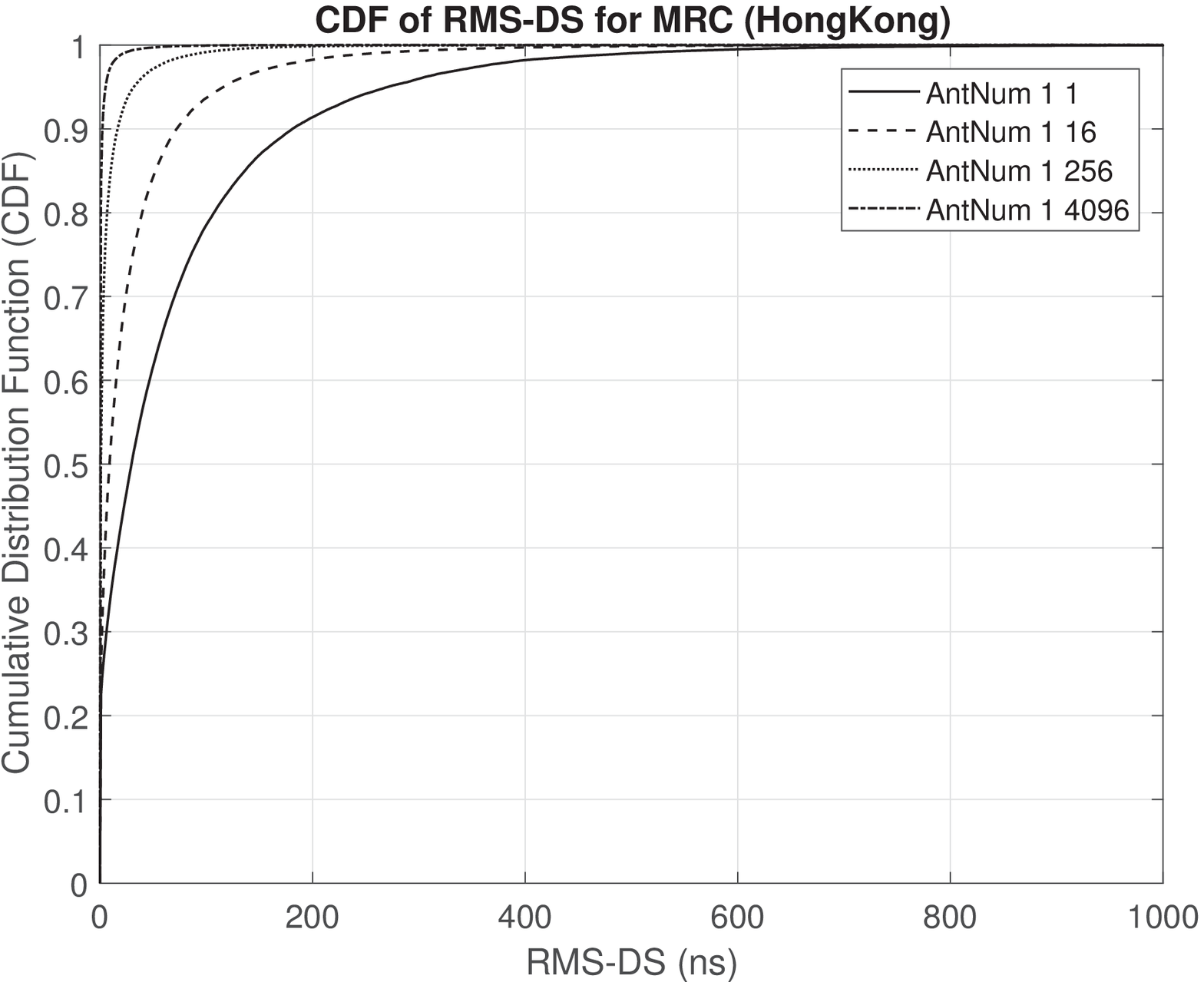}{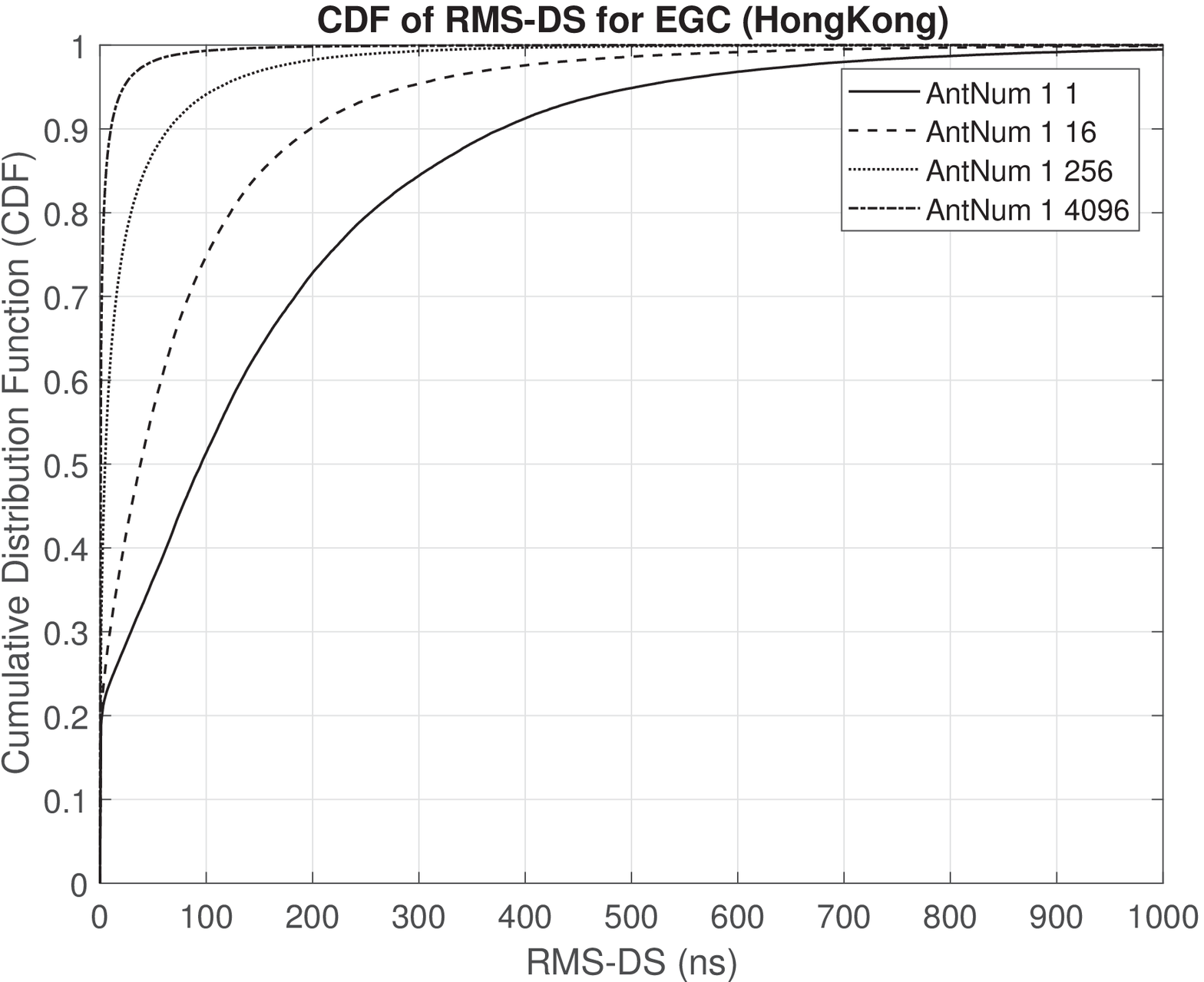}{\label{fig.HongKongCDF}}
{CDF of the RMS-DS for Hong Kong: (a) MRC or (b) EGC.}

\inserttwofigsV {0}{7.5cm}{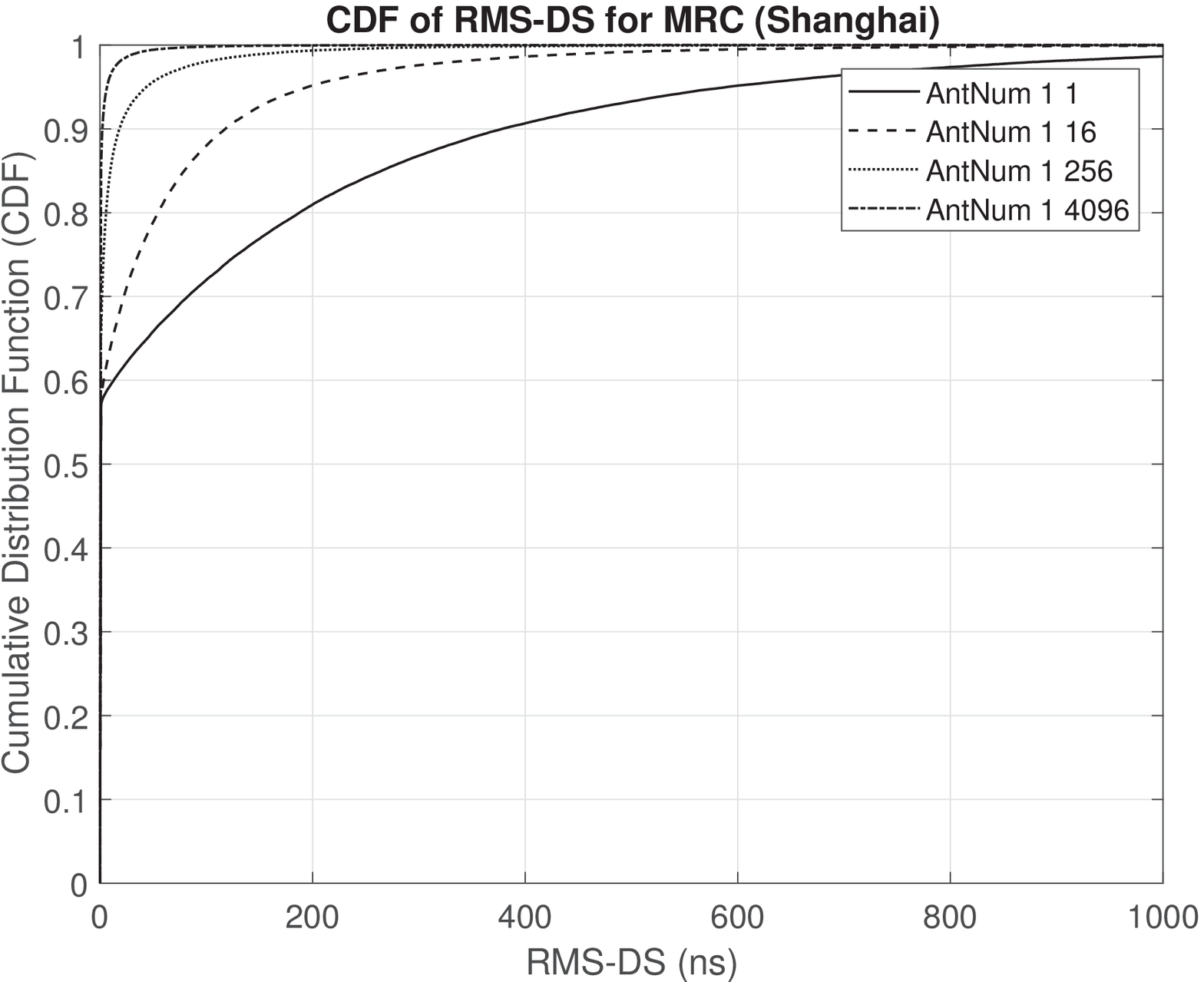}{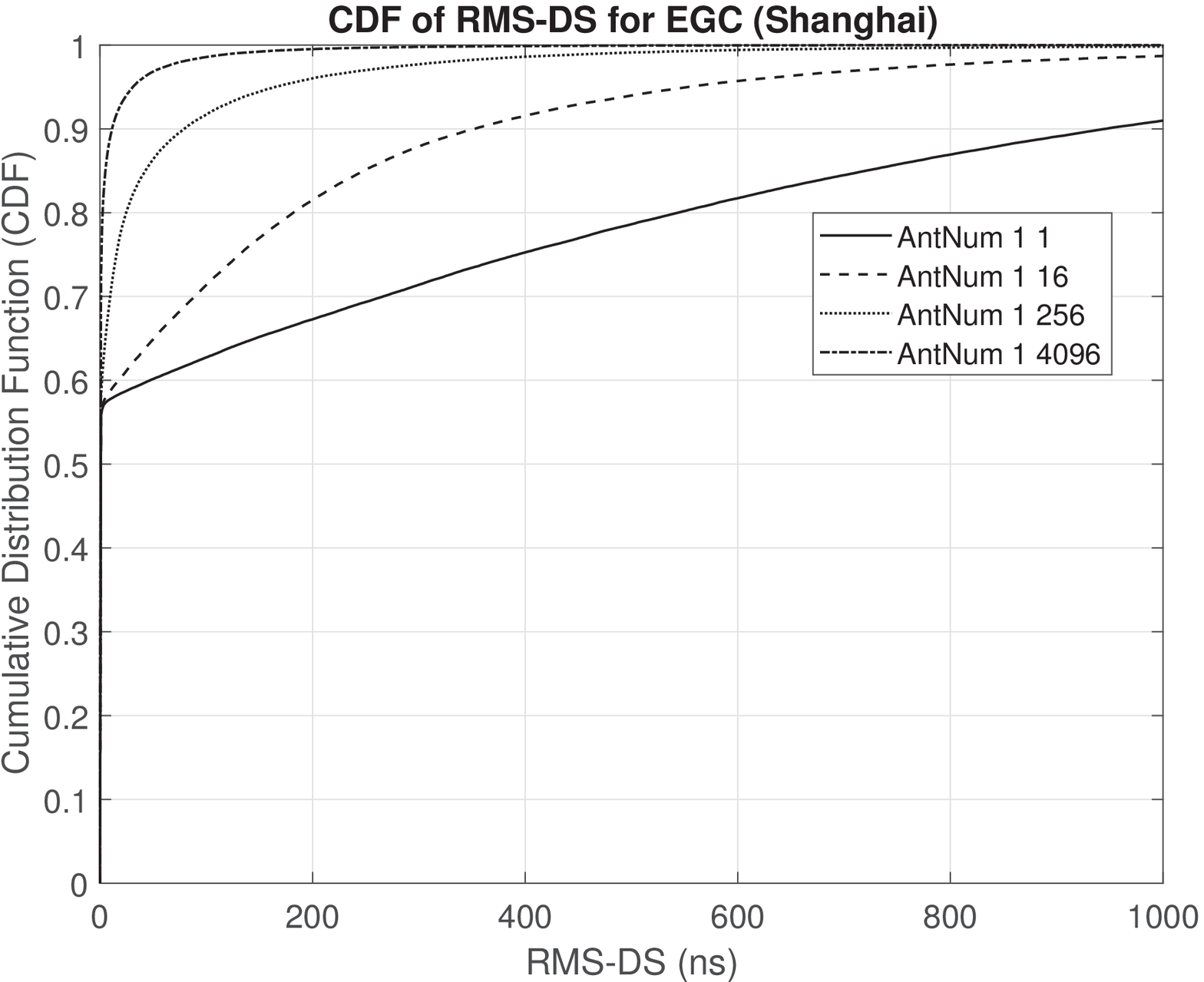}{\label{fig.ShanghaiCDF}}
{CDF of the RMS-DS for Shanghai: (a) MRC or (b) EGC.}

\appendices

\begin{figure*}[!t]
\normalsize
\setcounter{MYcounter}{\value{equation}}
\setcounter{equation}{19}
\begin{equation}\label{eq.SIMO-ISI-corrNUM}
    \bh^\herm[n] \ba(\alpha_{\textit{BS}}) = \sum_{j=1}^{P} g^*_{j,n}  \left( \sum_{m=0}^{M-1} e^{j2\pi \frac{d_{ant}}{\lambda} m [\cos(\alpha_{j})-\cos(\alpha_{\textit{BS}})]} \right)  =
    \begin{cases}
        M g_{k,n} & \text{, if $\alpha_{\textit{BS}}=\alpha_k$}
        \\
        \sum_{j=1}^{P} g_{j,n}
   \frac{1-e^{j2\pi \frac{d_{ant}}{\lambda} M[\cos(\alpha_{j})-\cos(\alpha_{\textit{BS}})]}}
  {1-e^{j2\pi \frac{d_{ant}}{\lambda}[\cos(\alpha_{j})-\cos(\alpha_{\textit{BS}})]}} & \text{, otherwise}
  \end{cases}
\end{equation}

\setcounter{equation}{20}

\begin{equation}\label{eq.SIMO-ISI-corrfin}
  \zeta^{(M)}_n =
        \begin{cases}
        \frac{g_{k,n}}{\sqrt{\sum_{i=1}^{P} | g_{i,n} |^2 +  \frac{1}{M}\sum_{i=1}^{P} \sum_{\substack{j=1 \\ j \neq i}}^{P} g^*_{i,n} g_{j,n}  \frac{1-e^{j2\pi \frac{d_{ant}}{\lambda} M[\cos(\alpha_{i})-\cos(\alpha_{j})]}}
  {1-e^{j2\pi \frac{d_{ant}}{\lambda}[\cos(\alpha_{i})-\cos(\alpha_{j})]}}}}   & \text{, if $\alpha_{\textit{BS}}=\alpha_k$}    \\

  \frac{\frac{1}{M}\sum_{j=1}^{P} g_{j,n}
   \frac{1-e^{j2\pi \frac{d_{ant}}{\lambda} M[\cos(\alpha_{j})-\cos(\alpha_{\textit{BS}})]}}
  {1-e^{j2\pi \frac{d_{ant}}{\lambda}[\cos(\alpha_{j})-\cos(\alpha_{\textit{BS}})]}}}{\sqrt{\sum_{i=1}^{P} | g_{i,n} |^2 +  \frac{1}{M}\sum_{i=1}^{P} \sum_{\substack{j=1 \\ j \neq i}}^{P} g^*_{i,n} g_{j,n}  \frac{1-e^{j2\pi \frac{d_{ant}}{\lambda} M[\cos(\alpha_{i})-\cos(\alpha_{j})]}}
  {1-e^{j2\pi \frac{d_{ant}}{\lambda}[\cos(\alpha_{i})-\cos(\alpha_{j})]}}}}

        & \text{, otherwise}
        \end{cases}
\end{equation}
\setcounter{equation}{\value{MYcounter}}
\hrulefill
\end{figure*}

\section{Wideband Channel Model} \label{App.WidebandChannel}
A spatial physical vector model for frequency-selective channels between a single antennas node and a Base Station equipped with $M$ antennas (ULA case) is derived here. Starting from the results in \cite{SalzWinters1994,NaguibPaulraj1996PerformanceWirelessCDMA}, the model derived here also includes the effects of transmit/receive pulse shaping filters which introduce correlation between the channel taps of the equivalent baseband, low-pass filtered, and baud-rate sampled channel impulse response.

The frequency and time selective channel between a single antennas UE and the $m$-th BS antenna can be written as follows ($m$=0,1,...,$M-1$):
\begin{equation}\label{eq.APP1}
  h^{(m)}(t,\tau) = \sum_{k=1}^{P} c_k(t) e^{-j2 \pi m \cos(\alpha_k(t))} \delta(\tau-\tau_k)
\end{equation}
where $c_k(t)$ time varying channel gain of the $k$-th path, $\alpha_k(t)$ time varying AoA of $k$-th path, $\tau_k$ time of arrival of $k$-th path, and $P$ is number of multipath components. Using vector notation, we can re-write \eqref{eq.APP1} as:
\begin{equation}\label{eq.APP2}
  \bh(t,\tau) = \sum_{k=1}^{P} c_k(t) \delta(\tau-\tau_k) \ba(\alpha_k(t))
\end{equation}
where $\bh(t,\tau)$ the $\medmuskip=0mu M\times 1$ is the vector collecting the impulse responses of all $M$ antennas, and vector $\ba(\alpha_k)=[1,\ e^{-j2 \pi \frac{d_{ant}}{\lambda} cos(\alpha_k(t))},\ \ldots,\ e^{-j 2 \pi (M-1) \frac{d_{ant}}{\lambda} cos(\alpha_k(t))}]^T$ is the steering vector.

The equivalent baseband, low-pass filtered, and baud-rate sampled $\medmuskip=0mu M\times 1$ vector channel impulse response can then be expressed as:
\begin{equation} \label{eq.APP3}
\bh_{eq}[k,n]\!=\!\iint \bh(kT_s-\xi,\tau) p_{t}(nT_s-\tau-\xi) p_{r}(\xi) d\tau d\xi,
\end{equation}
\noi where $p_{t}(t)$ is the transmit pulse shaping filter, $p_{r}(t)$ is the receive filter, and  $\tilde{p}(t)=p_{t}(t) \ast p_{r}(t)$ is a Raised Cosine filter.

Let us now assume that the channel is time-flat at least within the transmission of a frame or an OFDM symbol: $c_k(t)\approx c_k$ and $\alpha_k(t)\approx\alpha_k$. Then, we have also that $\bh_{eq}[k,n]\approx \bh_{eq}[n]$ holds. Under this assumption, we can simplify the double integral \eqref{eq.APP3} as follows ($n=0,1,\ldots,L-1$):
\begin{eqnarray} \label{eq.APP4}
  \!\!\!\!\! \bh_{eq}[n]  \!\!\!\!\! &=& \!\!\!\!  \sum_{k=1}^{P}  c_k \ba(\alpha_k) \int_{-\infty}^{\infty}  p_{t}(nT_S-\tau_k-\xi) p_{r}(\xi) d\xi \nonumber \\
    &=& \!\!\!\! \sum_{k=1}^{P}  c_k \ba(\alpha_k) \tilde{p}(t-\tau_k)\Bigr|_{t=nT_s} \!= \! \sum_{k=1}^{P}  g_{k,n} \ba(\alpha_k),
\end{eqnarray}
\noi where we have posed $g_{k,n}=c_k \tilde{p}(t-\tau_k)\Bigr|_{t=nT_S}$.

Finally, we can write the equivalent baseband, low-pass filtered, and baud-rate sampled $\medmuskip=0mu M\times 1$ vector SIMO time-flat and frequency-selective fading wideband channel impulse response of duration $L$ ($0\leq n \leq L-1$):
\begin{equation} \label{eq.APPfinal}
  \bh_{eq}[n] = \sum_{k=1}^{P}  g_{k,n} \ba(\alpha_k) = \bA \bg_{n}
\end{equation}
where $\bA=[\ba(\alpha_1), \ldots, \ba(\alpha_P)]$ and $\bg_{n}=[g_{1,n}, \ldots, g_{P,n}]^T$. The $\medmuskip=0mu M\times P$ matrix $\bA$ is a function of the number of antennas $M$ and the number of multipath components $P$, while the $\medmuskip=0mu P\times 1$ vector $\bg_{n}$ is a function of the gain and delay of all paths, the sampling instant $n$, and the transmit/receive filter cascade.

When transmit/receive filters are taken into account, the shape of $\tilde{p}(t)$ contributes to the observed ISI. Given $P>1$, $L$ can be smaller or greater than $P$ and even equal to $1$ (flat fading). It depends on how many resolvable paths there are, the associated path delays $\tau_k$, and the sampling rate $T_S$. If $P=1$ or if the delays $\tau_k$ are all very close to $\tau_1$, then the model derived here simplifies to a narrowband frequency-flat model such as the one reported in \cite{TsaiBuehrer2002ImpactAOAenergy}. In this case, we have $g_{k,n}=c_k$ at $n=0$ and $g_{k,n}=0$ at other sampling instants because $\tilde{p}(t)$ has a Raised Cosine characteristic.

For low values of $P$, the wideband model \eqref{eq.APPfinal} holds for a scenario with little or moderate scattering. However, for very large values of $P$, the model becomes equivalent to the usual WSSUS model for \emph{rich scattering} where all the entries of the $\medmuskip=0mu M\times L$ channel matrix $\bH=[\bh[0],\bh[1],\ldots,\bh[L-1]]$ become i.i.d. zero mean, complex Gaussian random variables.


\section{Proof of eq. \eqref{eq.SIMO-ISI-rholimit}} \label{App.Proof}
We can write the norm of $\bh[n]$ as:
\begin{IEEEeqnarray}{rCl}\label{eq.SIMO-ISI-corrDEN1}
  \parallel\bh[n]\parallel^2 &=& \bg^\herm_n \bA^\herm \bA \bg_n = M \sum_{i=1}^{P} | g_{i,n} |^2 \\
  &~& + \sum_{i=1}^{P} \sum_{\substack{j=0 \\ j \neq i}}^{P} g^*_{i,n} g_{j,n}  \frac{1-e^{j2\pi \frac{d_{ant}}{\lambda} M[\cos(\alpha_{i})-\cos(\alpha_{j})]}}
  {1-e^{j2\pi \frac{d_{ant}}{\lambda}[\cos(\alpha_{i})-\cos(\alpha_{j})]}} \nonumber
\end{IEEEeqnarray}
while we simply have $\parallel \ba(\alpha_{\textit{BS}}) \parallel^2 = \ba^\herm(\alpha_{\textit{BS}}) \ba(\alpha_{\textit{BS}}) = M$.

The numerator of $\zeta^{(M)}_n$ in eq. \eqref{eq.SIMO-ISI-corr} can be written as in eq. \eqref{eq.SIMO-ISI-corrNUM} and the value of $\zeta^{(M)}_n$ is finally given by eq. \eqref{eq.SIMO-ISI-corrfin}, both at the top of the page. If angles $\alpha_{i}$ and $\alpha_{j}$ in eq. \eqref{eq.SIMO-ISI-corrfin} are always different, then the ratio $\frac{1-e^{j2\pi M\textit{\ldots}}}{1-e^{j2\pi\textit{\ldots}}}$ is  bounded for any $M$. It is then immediate to recognize that taking the limit of eq. \eqref{eq.SIMO-ISI-corrfin} allows us to obtain eq. \eqref{eq.SIMO-ISI-rholimit} as $g_{k,n}=c_k$ for $n=0$ because only the $k$-th path is observed and $\tilde{p}(t)$ has a Raised Cosine characteristic (see the definition of $g_{k,n}$ in \eqref{eq.APP4}).


\balance

\bibliographystyle{IEEEtran}
\bibliography{IEEEabrv,All_Papers}

\end{document}

%% file: MyMacros.tex
\DeclareMathOperator{\tr}{\mathrm{tr}}

\DeclareMathOperator*{\argmin}{arg\,min}
\DeclareMathOperator*{\argmax}{arg\,max}

\newfont{\bbb}{msbm10 scaled 700}

\newfont{\set}{msbm10 scaled 1100}

\newcommand{\ba}{\mathbf{a}}
\newcommand{\bA}{\mathbf{A}}

\newcommand{\bg}{\mathbf{g}}

\newcommand{\bh}{\mathbf{h}}
\newcommand{\bH}{\mathbf{H}}

\newcommand{\bM}{\mathbf{M}}
\newcommand{\bn}{\mathbf{n}}

\newcommand{\br}{\mathbf{r}}
\newcommand{\bR}{\mathbf{R}}
\newcommand{\bs}{\mathbf{s}}

\newcommand{\bv}{\mathbf{v}}

\newcommand{\bx}{\mathbf{x}}

\newcommand{\by}{\mathbf{y}}

\newcommand{\Psim}{\hbox{\boldmath$\Psi$}}

\newcommand{\Xim}{\hbox{\boldmath$\Xi$}}

\newcommand{\diag}{{\hbox{diag}}}

\newcommand{\herm}{{\sf H}}

\newcommand{\noi}{\noindent}

\definecolor{myRED}{RGB}{255, 0, 0}
\definecolor{myBLUE}{RGB}{0,0,255}
